\documentclass[%
 reprint,
 amsmath,amssymb,superscriptaddress,floatfix,
prl,
]{revtex4-2}
\usepackage{CJK}
\usepackage{amsmath}
\usepackage{mathtools}
\usepackage{graphicx}  
\usepackage{physics}
\usepackage{graphicx}
\usepackage{dcolumn}
\usepackage{bm}
\usepackage{comment}
\usepackage{amssymb}
\usepackage{epstopdf}
\usepackage{inputenc}
\usepackage{esint}

\usepackage{autobreak}

\begin{document}

\title{On the nature of optical thermodynamic pressure exerted in highly multimoded nonlinear systems}

\author{Huizhong Ren}
\affiliation{Ming Hsieh Department of Electrical and Computer Engineering, University of Southern California, Los Angeles, California 90089, USA}
\author{Georgios G. Pyrialakos}
\affiliation{Ming Hsieh Department of Electrical and Computer Engineering, University of Southern California, Los Angeles, California 90089, USA}
\affiliation{CREOL, The College of Optics and Photonics, University of Central Florida, Orlando, Florida 32816, USA}

\author{Fan O. Wu}%
\affiliation{CREOL, The College of Optics and Photonics, University of Central Florida, Orlando, Florida 32816, USA}
\author{Pawel S. Jung}%
\affiliation{CREOL, The College of Optics and Photonics, University of Central Florida, Orlando, Florida 32816, USA}
\affiliation{Faculty of Physics, Warsaw University of Technology, Koszykowa 75, 00-662, Warsaw, Poland}

\author{Nikolaos K. Efremidis}%
\affiliation{Department of Mathematics and Applied Mathematics, University of Crete, Heraklion, Crete 70013, Greece}

\author{Mercedeh Khajavikhan}%
\affiliation{Ming Hsieh Department of Electrical and Computer Engineering, University of Southern California, Los Angeles, California 90089, USA}

\author{Demetrios Christodoulides}%
\email{demetri@usc.edu}
\affiliation{Ming Hsieh Department of Electrical and Computer Engineering, University of Southern California, Los Angeles, California 90089, USA}

\begin{abstract}
The theory of optical thermodynamics provides a comprehensive framework that enables a self-consistent description of the intricate dynamics of nonlinear multimode photonic systems. This theory, among others, predicts a pressure-like intensive quantity ($\hat{p}$) that is conjugate to the system's total number of modes ($M$) - its corresponding extensive variable. Yet at this point, the nature of this intensive quantity is still nebulous. In this paper, we shed light on the physical origin of the optical thermodynamic pressure and demonstrate
its dual essence. In this context, we rigorously derive an expression that splits $\hat{p}$ into two distinct components, a term that is explicitly tied to the electrodynamic radiation pressure and a second entropic part that responsible for the entropy change in an irreversible expansion process. We utilize this result to establish a formalism that simplifies the quantification of radiation pressure under nonlinear equilibrium conditions, thus eliminating the need for a tedious evaluation of the Maxwell stress tensor. Our theoretical analysis is corroborated by numerical simulations carried out in highly multimoded nonlinear optical structures. These results may provide a novel way in predicting and controlling radiation pressure processes in a variety of nonlinear electromagnetic settings.

\end{abstract}

\maketitle

\emph{Introduction.}\textemdash The presence of nonlinearity in highly multimoded photonic systems can give rise to a broad array of novel phenomena \cite{renninger2013optical,wright2015controllable,wright2017spatiotemporal,longhi2003modulational,eisenberg1998discrete,petersen2014mid,pourbeyram2013stimulated,zitelli2022characterization,PhysRevLett.116.183901,gao2023all,mao2021synchronized,ahsan2018graded,sun2022multimode,pourbeyram2022direct,marques2023observation,PhysRevLett.130.063801} but at the same time introduces several new fundamental theoretical challenges in understanding and predicting the intricate spatio-temporal dynamics that emerge \cite{poletti2008description,mafi2012pulse,agrawal2000nonlinear}. To address many of these issues, quite recently, an optical thermodynamic theory was put forward \cite{wu2019thermodynamic,makris2020statistical,parto2019thermodynamic}, capable of describing the collective behavior of the ``photon gas" in heavily multimoded systems under weak nonlinearity (Fig.1 (a)). In this formalism, an extensive optical entropy was self-consistently introduced $S=S(U,M,\mathcal{P})$ that directly involves the two invariants of the system, i.e., the ``internal energy" $U$ and total optical power $\mathcal{P}$ as well as the total number of modes $M$. In general, it was found that at equilibrium, the power distribution among the various modes obeys a Rayleigh-Jeans law \cite{pourbeyram2022direct,marques2023observation,PhysRevLett.130.063801} - a response that in turn maximizes the optical entropy \cite{wu2019thermodynamic,makris2020statistical}. In this regard, the theory of optical thermodynamics \cite{wu2019thermodynamic,makris2020statistical,parto2019thermodynamic} has provided a clear pathway in harvesting notions from statistical mechanics when studying a variety of nonlinear multimode optical configurations, like optical fibers, waveguide multicore lattices, and cavity systems. \cite{pourbeyram2022direct,marques2023observation,PhysRevLett.130.063801,wu2019thermodynamic,makris2020statistical,parto2019thermodynamic,mangini2022statistical,PhysRevA.103.043517,podivilov2022thermalization,PhysRevLett.128.123901,PhysRevX.10.031024,shi2021controlling}.   
 
 A fundamental tenet of statistical mechanics states that, for each extensive variable, there should exist a conjugate intensive quantity that acts as a thermodynamic force \cite{pathria2016statistical}. In the context of optical thermodynamics, the optical temperature and chemical potential can be defined through the fundamental thermodynamic equation, according to $1/T=\partial S/\partial U$ and $\mu/T=-\partial S/\partial\mathcal{P}$, respectively \cite{wu2019thermodynamic}. Most importantly, these two intensive quantities $(T,\mu)$ directly govern the energy and power exchange $(\Delta U,\Delta\mathcal{P})$ between optical subsystems in thermal contact with each other \cite{wu2019thermodynamic}, in analogy to classical thermodynamics. At the same time, the fundamental equation of thermodynamics predicts an additional intensive quantity, the so called ``optical thermodynamic pressure", i.e., $\hat{p}/T=\partial S/\partial M$ that, in principle, results from a change in the total number of modes $M$. However, the physical manifestations of this thermodynamic quantity $\hat{p}$ have so far remained elusive. If indeed it represents a thermodynamic force, could a possible interpretation of $\hat{p}$ be tied to actual pressure effects, such as those resulting from radiation pressure forces at the boundaries of dielectric materials?

\begin{figure}
\centering
\includegraphics[scale=0.44]{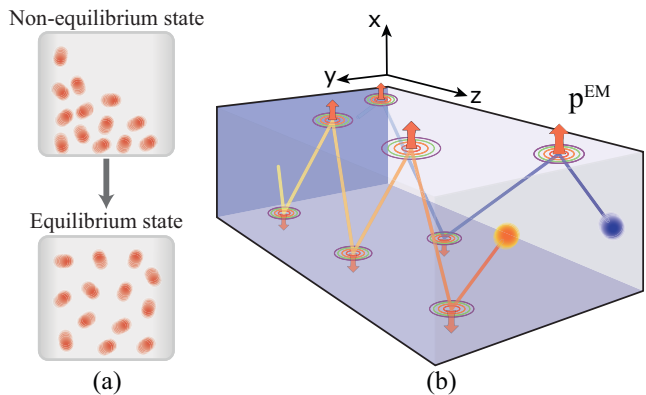}
\caption{\label{fig1}  (a). Photons in a nonlinear multimode optical system. An optical arrangement can be excited with a random initial state. After the modes are thermalized, an equilibrium state is attained by maximizing the optical entropy. (b). Different photon modes can bounce at the waveguide boundaries with distinct transverse wave vectors, resulting in unique radiation pressure contributions. At equilibrium, the total pressure can be deduced by summing incoherently the contributions from all modes once weighted with respect to a Rayleigh-Jeans distribution.}
\end{figure}

Thus far, radiation pressure forces have found significant applications in laser cooling \cite{RevModPhys.58.699,phillips1998nobel,ashkin1978trapping}, optical trapping and manipulation of microscopic objects \cite{ashkin1970acceleration,ashkin1985observation,yang2009optical,schmidt2007optofluidic,soltani2014nanophotonic}. At the same time, opto-mechanical arrangements based on strong phonon-photon coupling effects are nowadays actively pursued \cite{RevModPhys.86.1391,kippenberg2008cavity,PhysRevX.2.011008,povinelli2005evanescent,rakich2007trapping,li2008harnessing,wiederhecker2009controlling,manipatruni2009optical}. One way to understand these phenomena is through the perspective of momentum exchange between an object/medium and the electromagnetic field. For example, as Fig.1 (b) shows, in a waveguide system, the bouncing photons conveyed in each guided mode continuously exert an electromagnetic radiation pressure perpendicular to the boundaries, which can be determined via the Maxwell stress tensor \cite{zangwill2013modern,stratton2007electromagnetic}.
In general, the Maxwell stress tensor can be employed when the field distribution is known a priori, such as in a single mode linear waveguide. However, standard electromagnetic techniques fail in chaotic systems (such as multimoded nonlinear configurations) where, at any given instance of time, the system exists in an random and unknown state. To tackle such a problem, it would be necessary to employ the tool-set of thermodynamics and statistical mechanics. In a statistical mechanics context, historically, electromagnetic pressure has been mainly studied in black-body systems under thermal equilibrium conditions \cite{kelly1981thermodynamics,leff2002teaching,sonnleitner2013attractive,haslinger2018attractive,yu2013enhancing,zhu2015radiative}.  An interesting question to ask is whether a connection between radiation pressure and statistical mechanics can be constructed for purely optical environments where thermalization ensues because of ``photon-photon collisions", i.e., without the intervention of other quasiparticles like phonons. 

In this manuscript, we elucidate the physical nature of the optical thermodynamic pressure, as defined within the context of optical thermodynamics, and formally derive an equation that unifies it with the electromagnetic radiation pressure. This allows us to rigorously associate the average radiation pressure of large chaotic systems with thermodynamic quantities such as the optical temperature and chemical potential, deriving a universal thermodynamic equation. Through this equation, the radiation pressure at thermal equilibrium can be calculated in an effortless manner. Our results are found in agreement with the Maxwell stress tensor formalism that can be exclusively evaluated through extensive numerical simulations. As a result, it is shown that optical thermodynamics \cite{wu2019thermodynamic,makris2020statistical,parto2019thermodynamic} can directly provide the radiation pressure resulting from thousands of modes (once optical thermal equilibrium is reached) without any knowledge of the vectorial structure of the electromagnetic modes involved.

\emph{Background.}\textemdash We begin this work by considering an arbitrary nonlinear multimode optical waveguide system supporting a finite number of $M$ bound modes, propagating along the axial direction $z$. Each mode $|\psi_k\rangle$ and its corresponding propagation constant $\beta_k$ can be obtained from the pertinent eigenvalue problem associated with the evolution equation $i d|\Psi\rangle/dz=-\hat{H_L}|\Psi\rangle$ where $\hat{H_L}$ is the linear Hamiltonian operator of the optical system itself. Under weakly nonlinear conditions, the total Hamiltonian of this arrangement is primarily dominated by its linear component, i.e., $H\approx H_L$ \cite{wu2019thermodynamic,makris2020statistical}. In this regime, the role of nonlinearity is merely to ergodically and chaotically promote power exchange among the various states or modes - in a way akin to particle collisions in ideal gases responsible for thermalization.

In conservative systems, one can directly identify two invariants. These are: (i) the ``internal energy" of the system $U=-\sum_{k=1}^M{\beta_kn_k}$ that physically represents the Minkowski longitudinal electrodynamic momentum flow density \cite{haus1976electromagnetic} and (ii) the total optical power $\mathcal{P}=\sum_{k=1}^M{n_k}$ transported in this guiding configuration. In the above expressions, $n_k$ represents a quantity that is proportional to the average power $|c_k|^2=\mathcal{P}_0n_k$ conveyed by mode $|\psi_k\rangle$ where the arbitrary level $\mathcal{P}_0$ denotes the power of each discrete power-packet (see Supplementary). Note that $c_k(z) = \langle\psi_k|\Psi(z)\rangle$.  In all cases, these two independent conserved quantities are completely specified by the initial excitation conditions \cite{parto2019thermodynamic}. As indicated in previous studies, upon thermalization, the expectation value of the power occupancy associated with mode $k$, eventually settles into a Rayleigh-Jeans (RJ) distribution, that is, $n_k=-T/(\beta_k+\mu)$ where $T$ represents the optical temperature of the diluted optical gas and $\mu$ is the corresponding chemical potential \cite{wu2019thermodynamic,makris2020statistical,parto2019thermodynamic,picozzi2014optical}. Interestingly, these systems display the following global equation of state $U-\mu\mathcal{P}=MT$ (see Supplementary).

\emph{Optical thermodynamic pressure.}\textemdash Of importance is the classical entropy associated with these photonic multimode systems \cite{wu2019thermodynamic,makris2020statistical,parto2019thermodynamic,mangini2022statistical,PhysRevA.103.043517,podivilov2022thermalization,PhysRevLett.128.123901,PhysRevX.10.031024,shi2021controlling}
\begin{equation}
S=\sum_{k=1}^{M}\ln{n_k}.
\end{equation}
In the microcanonical ensemble, the entropy is expressed as a function of the other three extensive variables, $S=S(U,M,\mathcal{P})$, a relation that directly leads to the fundamental thermodynamic equation $TdS=dU-\mu d\mathcal{P}+\hat{p}dM$  where $1/T = \partial S / \partial U$, $\mu/T = -\partial S / \partial \mathcal{P}$ and 
\begin{equation}
\frac{\hat{p}}{T}=\frac{\partial S}{\partial M}
\end{equation}
Here $\hat{p}$ is what we call ``optical thermodynamic pressure”. Like the other two intensive quantities $T$ and $\mu$ that are conjugate to their corresponding variables $U$ and $\mathcal{P}$, the optical thermodynamic pressure $\hat{p}$ mathematically emerges from its complementarity with respect to the number of modes $M$. While $T$ and $\mu$ act in every respect as thermodynamic forces that govern the exchange of $U$ and $\mathcal{P}$ \cite{wu2019thermodynamic}, at this point it is unknown if indeed a similar physical significance can be ascribed to $\hat{p}$ - the topic of this paper. To address this fundamental issue we consider the optical thermodynamic pressure as obtained from the entropy at thermal equilibrium 
\begin{equation}
    \frac{\hat{p}}{T}=\left.\frac{\partial}{\partial M}\right\vert_{\mathcal{P},U}\left\{\sum_{k=1}^{M}\ln{\left[-\frac{T}{\beta_{k}(M)+\frac{1}{\mathcal{P}}(U-MT)}\right]}\right\},
\end{equation}
where in obtaining Eq. (3) we made use of Eqs. (1-2) in conjunction with the RJ distribution and the global equation of state (see Supplementary). In Eq. (3), $\beta_k(M)$ denotes the propagation eigenvalue constant of mode $k$ which of course is also a function of $M$.  These eigenvalues are arranged according to $\beta_1\geq\beta_2\geq...\geq\beta_M$, where $\beta_1$ stands for the eigenvalue of the ground state while $\beta_M$ corresponds to that associated with the highest-order mode supported in this system \cite{okamoto2021fundamentals}. The following relation proves useful in our analysis 
\begin{equation}
    \left.\frac{\partial S(U,M,\mathcal{P};T)}{\partial M}\right\vert_{\mathcal{P},U}=\frac{\partial S}{\partial M}+\frac{\partial S}{\partial T}\frac{\partial T}{\partial M}=\frac{\partial S}{\partial M}.
\end{equation}
In obtaining Eq. (4) we used the fact that at thermal equilibrium, $\partial S/\partial T=0$ as shown in the Supplementary. Given that the system is highly multimoded, the summation in Eq. (3) can be written as an integral (a first order approximation to the Euler–Maclaurin formula \cite{ARFKEN2013551})
\begin{equation}
    \frac{\hat{p}}{T}=\left.\frac{\partial}{\partial M}\right\vert_{\mathcal{P},U}\left\{\int_{1}^{M}{\ln{\left[-\frac{T}{\beta(M,k)+\frac{1}{\mathcal{P}}(U-MT)}\right]}dk}\right\}.
\end{equation}
The partial derivative in Eq. (5) can now be performed by employing the Leibniz integral rule (see Supplementary), and yields 
\begin{equation}
    \frac{\hat{p}}{T}=\frac{1}{T}\int_{1}^{M}{n_k\frac{\partial \beta(M,k)}{\partial M}dk}+\left[\ln{\left(-\frac{T}{\beta_M+\mu}\right)}-1\right].
\end{equation}

Equation (6) provides a general expression for the optical thermodynamic pressure. To understand the physical ramifications of the terms involved in Eq. (6), let us first recall that the Minkowski longitudinal electrodynamic momentum flow density $\mathcal{M}$ \cite{haus1976electromagnetic} is given by $\mathcal{M}\propto-U=\sum_{k=1}^M{\beta_kn_k}$.  In this respect, the first term in Eq. (6) corresponds to a change in the longitudinal momentum flow when the supported number of modes $M$ varies adiabatically. Typically, for a waveguide structure, a change in the number of modes can be achieved by increasing the structure size, in other words, “expanding” the waveguide structure.  From previous studies \cite{povinelli2005evanescent,povinelli2005high,ma2011mechanical,rakich2011scaling,rakich2009general,rakich2010tailoring,rodrigues2017optical,rodrigues2017rigorous,rodrigues2019geometric,miri2018optical,ren2022rigorous}, one finds that the local pressure $p^{EM}_k$ exerted by mode $k$ (carrying power $P_k=\mathcal{P}_0n_k$) can be directly related to the total work performed during a virtual expansion through  $\int_{\mathcal{C}}p^{EM}_k dl=\omega^{-1}P_k d\beta_{k}/d\xi$  where $\omega$ is the angular frequency of the optical field, $\xi$ represents the dimension relevant to the waveguide expansion and the line integral is performed around the waveguide boundary ($\mathcal{C}$) in the $x-y$ plane. In a waveguide arrangement, the local pressure $p^{EM}_k$ depends on the electromagnetic field profile and hence it may not be uniform around the circumference. In this respect, one can define a spatially averaged electrodynamic pressure $\bar{p}^{EM}_{k}$, according to $\bar{p}^{EM}_{k}=W^{-1}\int_{\mathcal{C}}p^{EM}_k dl$ where $W$ is the relevant boundary length and $p^{EM}_k$ is the local pressure at each point at the boundary. For example, in a regular optical fiber with radius $a$, the electromagnetic force tends to expand the fiber along the radial direction. In this case, the line integral will be performed around the fiber circumference and hence $W=2\pi a$. By considering that the total number of supported modes is related to $\xi$ through $M=M(\xi)$, the average electrodynamic pressure produced by each mode can be rewritten as $\bar{p}^{EM}_{k}=(\omega W)^{-1}P_k (dM/d\xi) (d\beta_k/dM)$. Given that upon optical thermalization, the modal fields happen to be mutually incoherent with respect to each other, one can then obtain the total average electrodynamic pressure ${p}_T^{EM}$ by summing the contributions from all the modes, i.e., ${p}_T^{EM}=\sum_{k}{\bar{p}^{EM}_{k}}$.  From all the above considerations, it is straightforward to show that the first term in the RHS of Eq. (6) represents the total average electrodynamic pressure and hence Eq. (6) can now be recast as 
\begin{equation}
    \hat{p}=Q\cdot{p}_T^{EM}+T\ln\left({-\frac{T}{\beta_M+\mu}}\right)-T,
\end{equation}
where the prefactor $Q=\omega W\mathcal{P}_0^{-1}/(dM/d\xi)$ is completely determined by the waveguide system itself  (see Supplementary S5).

Equation (7) indicates that the thermodynamic pressure in highly multimode systems is directly related to the electrodynamic pressure once a universal entropic term that solely depends on $T$ and $\mu$ is taken into account. To elucidate the physical significance of this additional universal term $T\ln\left[-T/(\beta_M+\mu)\right]-T$, we make use of the fundamental thermodynamic equation $TdS=dU-\mu d\mathcal{P}+\hat{p}dM$. If the total number of modes $M$ is increased by $dM$, after the waveguide arrangement has virtually expanded, both the entropy and the electromagnetic momentum change by an amount ($dS,dU$) while the power flowing in this system $\mathcal{P}$ remains invariant, i.e., $d\mathcal{P}=0$. In this regard, the fundamental thermodynamic equation is reduced to $TdS=dU+\hat{p}dM$. From our previous analysis one can readily show that $-dU/dM=\int_{1}^{M}{n_k\left[\partial \beta(M,k)/\partial M\right]dk}=Q\cdot p_T^{EM}$. As a result we find that the following relations will hold true for such an adiabatic irreversible expansion: $dS/dM = \ln\left[-T/(\beta_M+\mu)\right]-1$, $-dU/dM=Q\cdot{p}_T^{EM}$. These last two expressions imply that if the waveguide expands slowly enough, the universal term $T\ln\left[-T/(\beta_M+\mu)\right]-T$ in Eq. (7) will be responsible for the entropy increase while the the elctrodynamic pressure ${p}_T^{EM}$ will account for the resulting longitudinal momentum flow change. We must emphasize that the optical thermodynamic pressure always exists in a thermalizing electromagnetic system and does not require such an expansion process to manifest itself. Moreover, to illustrate how a lateral force can lead to a change in the longitudinal momentum flow we include a rigorous proof in Supplementary S9.

Of interest is to consider optical multimode systems where the density of states remains self-similar during expansion, a condition that is upheld by the majority of optical arrangements when dealing with a large number of modes. In this case, the thermodynamic Euler equation leads to the following simple expression \cite{wu2019thermodynamic,makris2020statistical}
\begin{equation}
    \frac{\hat{p}}{T}=\frac{S}{M}-1.
\end{equation}
From here, one quickly obtains the following relation
\begin{equation}
    p_T^{EM}=\frac{T}{Q}\left[\frac{S}{M}-\ln{\left(-\frac{T}{\beta_M+\mu}\right)}\right].
\end{equation} 
Equation (9) represents an important result: it allows one to obtain in an effortless manner the average optical pressure once a system attains a RJ thermal equilibrium at a temperature $T$ and chemical potential $\mu$. This can be accomplished by only knowing the linear spectrum of the system $\beta_n$ and the initial excitation conditions that specify the invariants $U,\mathcal{P}$. From this minimal information, one can directly calculate $S,T,\mu$ and therefore $p_T^{EM}$ from Eq. (9). This is quite impressive given that it is possible to obtain the average electrodynamic pressure without having any detailed knowledge as to the electromagnetic vectorial structure of the modes themselves - an ingredient traditionally required to compute the corresponding Maxwell stress tensor elements that lead to the optical pressure upon surface integration. These results can be further extended for non self-similar systems (upon expansion), as shown in the Supplementary.

\begin{figure}
\centering
\includegraphics[scale=0.45]{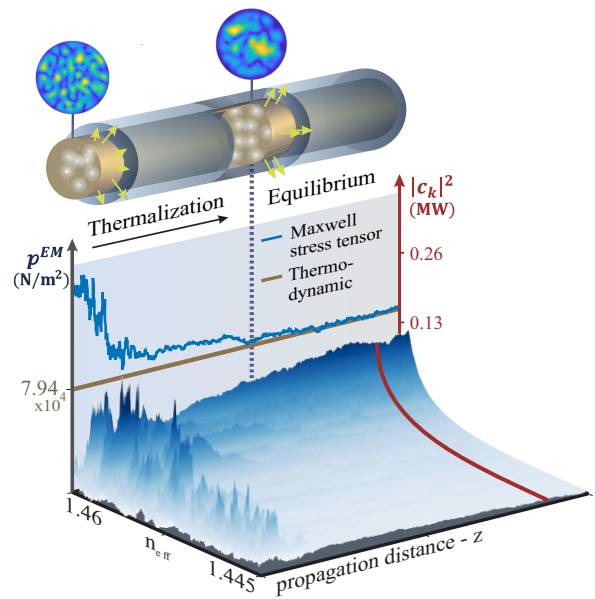}
\caption{\label{fig2} A beam with an irregular spatial profile is injected into a multi-mode weakly guided step-index fiber and undergoes nonlinear evolution, thermalizing into a RJ distribution at the output. Here, the power occupancies $|c_k|^2$ of all guided modes are shown as a two-dimensional surface plot, whose axes involve the effective refractive index of each mode ($n_{eff,k}=\beta_k/k_0$) and the propagation distance $z$. The electromagnetic pressure $p^{EM}$ is calculated at each $z$-point via the Maxwell stress tensor (blue curve), displaying a gradual relaxation towards a stable equilibrium value. The brown line corresponds to the electromagnetic pressure calculated via optical thermodynamics (Eq. (9)). The two curves are in excellent agreement after equilibrium is established. }
\end{figure}

\emph{Numerical results.}\textemdash To exemplify the validity of our theoretical results, we investigate electromagnetic pressure effects in a weakly guided step-index fiber with a radius $a=15 \mu m$, as shown in Fig. 2. The core refractive index of this fiber is $n_1=1.46$ while in the cladding is $n_2=1.4454$. This fiber supports $M=86$ modes \cite{okamoto2021fundamentals} in each polarization and conveys in total 2MW of power at a wavelength of $1064$nm. Our quasi-CW beam propagation simulations show that once the fiber is excited (at $z=0$) with an arbitrary spatial beam profile, it gradually undergoes thermalization into a RJ distribution-eventually displaying a beam self-cleaning effect at the output. Here, thermalization ensues due to small temporal fluctuations in the input power, which are emulated by introducing a $2\%$ power variation across 100 identical ensembles. At the output, nonlinearity results in maximum decoherence between the modes, a necessary factor in establishing a RJ distribution. The electromagnetic pressure at each point $z$ along the propagation axis can be monitored by performing a spatial averaging of the local forces per unit area along the circumference of the fiber, i.e., $p^{EM}=(2\pi)^{-1}\int_0^{2\pi}f(\theta)d\theta$, where $f(\theta)$ can be computed using the Maxwell stress tensor or the Minkowski-Helmholtz formula (see Supplementary). During thermalization, the total electromagnetic pressure - obtained by averaging over all ensembles - relaxes into an equilibrium value that remains constant throughout all subsequent observation distances. Alternatively, this equilibrium value can be evaluated effortlessly using Eq. (9) where for step-index fibers $Q=4\pi c\mathcal{P}_0^{-1}/[k_0(n_1^2-n_2^2)]$ (see Supplementary) and $\beta_M\simeq k_0n_2$ with $k_0=\omega/c $. The two results, i.e., the theoretical value predicted by Eq. (9), indicated by the brown line in Fig. 2, and the equilibrium value of the electromagnetic pressure as obtained from numerical simulations are in excellent agreement with each other, providing compelling evidence as to the power of the thermodynamic approach. It is worth noting that Eq. (9) involves solely thermodynamic variables that can be directly extracted from the optical beam profile at the input, in contrast to conventional approaches that require a full simulation of the nonlinear propagation dynamics, as demonstrated in this example.

\begin{figure}
\includegraphics[scale=0.8]{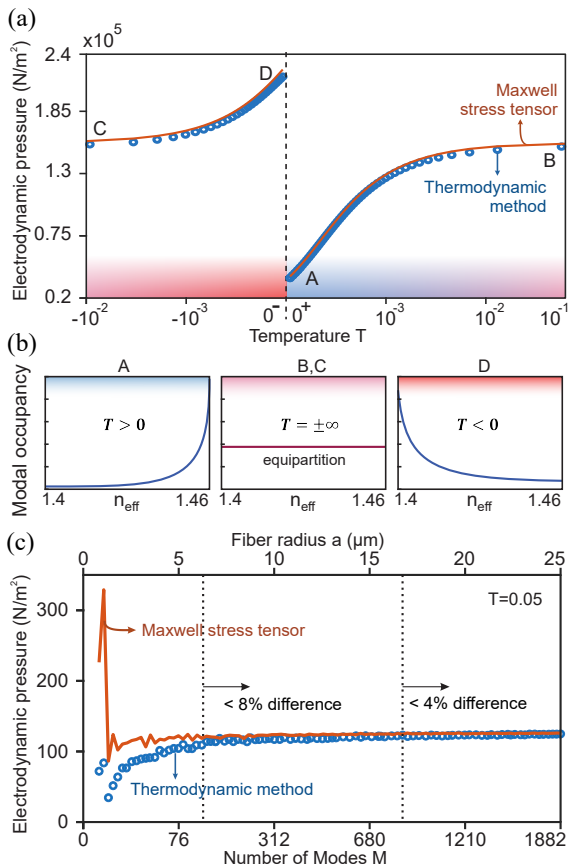}
\caption{\label{fig3}
(a). Electrodynamic pressure in a high contrast step-index fiber under different initial excitation conditions. The blue dots represent the pressure calculated via the optical thermodynamic method (Eq. (9)), while the red line depicts the pressure as obtained from the Maxwell stress tensor. (b). Modal occupancies corresponding to different temperature regimes. (c). Electrodynamic pressure exerted on the core-cladding interface of a high contrast step-index fiber as the radius $a$ increases. For $a>16
\mu$m, the error is less than $4\%$. Clearly, as the number of modes increases, the two approaches yield results that asymptotically converge to each other. In Figs. 3(a) and (b), $a=20\mu$m, $n_1=1.46, n_2=1.4$, $\lambda_0=1064$nm, and $P=2$MW. In 3(c) all structural parameters remain the same except that the core radius is allowed to vary.}
\end{figure}

To further demonstrate the universality of our methodology, we next evaluate the electrodynamic pressure for a high contrast step-index fiber with a radius $a=20 \mu m$. This is done across a continuous range of optical energies $U$, covering almost the entirety of accessible optical temperatures $[-\infty,+\infty]$. This fiber ($n_1=1.46$ and $n_2=1.4$) supports in total $M=1210$ modes (in both polarizations) at $\lambda_0=1064$nm \cite{okamoto2021fundamentals}, that include full vectorial TE, TM, EH and HE modes \cite{okamoto2021fundamentals}. In all cases, the total optical power is taken to be $P=2$MW. The results in Fig. 3(a) show that indeed these two methods are in excellent agreement across the entire temperature range. Note that for the thermodynamic method $Q=2\pi c\mathcal
{P}_0^{-1}/[k_0(n_1^2-n_2^2)]$ for a step-index fiber (see Supplementary). In Fig. 3(b), we plot the corresponding Rayleigh-Jeans distributions under positive and negative temperature conditions (points A and D in Fig. 3(a)) where the lower and higher-order modes are favored, respectively. Meanwhile, at infinite temperatures, $T\rightarrow\pm\infty$ (points B and C in Fig. 3(a)), power equipartition ensues among modes. Interestingly, as $T\rightarrow 0^-$, the electrodynamic pressure increases. Intuitively, this can be understood given that the predominant higher-order modes tend to bounce more frequently at the core-cladding interface - thus boosting up the pressure. The converse is true as $T\rightarrow 0^+$ in which case the lowest order modes (bouncing at very small angles) are occupied. In this case, the electrodynamic pressure is at its lowest, in accordance with Fig. 3(a). Note that caution should be taken in the limit $T\rightarrow 0^{\pm}$ where optical condensates can manifest themselves. In this extreme narrow regime, the thermodynamic analysis must be reconsidered given that the steps between Eqs. (3-6) should be reassessed. Finally, we investigate the electrodynamic pressure for varying fiber radii $a$ when the initial ($U,\mathcal{P}$) are such that the temperature and chemical potential ($T,\mu$) in the RJ distribution remain the same. As illustrated in Fig. 3(c), the discrepancy between the two methods diminishes as the number of modes increases, an expected outcome considering that the derivation of Eq. (9) relies on the assumption that the optical system is highly multimoded. 

\emph{Conclusion.}\textemdash In this Letter, we have investigated the physical essence of the optical thermodynamic pressure, as defined under the recently developed framework of optical thermodynamics. We formally demonstrated that this optical thermodynamic quantity is directly tied to the actual electromagnetic pressure once a universal entropic component is taken into account. This result provides an altogether new approach to compute in an effortless manner radiation pressure effects in highly multimoded nonlinear arrangements, without requiring any knowledge of the complex electromagnetic modal field distributions. We must emphasize that the theoretical analysis presented herein is general and can be applied to any electromagnetic system that can display thermalization. Of interest would be to experimentally observe these processes. In this regard,  radiation pressure induced liquid core fibers with high nonlinearities could be a promising direction \cite{PhysRevLett.101.014501,freysz1985giant}. 

 This work was partially supported by ONR MURI (N00014-20-1-2789), AFOSR MURI (FA9550-20-1-0322, FA9550-21-1-0202), DARPA (D18AP00058), Office of Naval Research (N00014-16-1-2640, N00014-18-1-2347, N00014-19-1-2052, N00014-20-1-2522, N00014-20-1-2789), National Science Foundation (NSF) (DMR-1420620, EECS-1711230, CBET 1805200, ECCS 2000538, ECCS 2011171), Air Force Office of Scientific Research (FA9550-14-1-0037,  FA9550-20-1-0322, FA9550-21-1-0202), MPS Simons collaboration (Simons grant 733682), W. M. Keck Foundation, USIsrael Binational Science Foundation (BSF: 2016381), US Air Force Research Laboratory (FA86511820019) and the Qatar National Research Fund (grant NPRP13S0121-200126). G.G.P. acknowledges the support of the Bodossaki Foundation. 

H.R and G.G.P contributed equally to this work.


\nocite{*}

\bibliography{reference}

\end{document}